\documentclass[12pt]{iopart}

\usepackage{graphicx}

\begin{document}
\title[Quantum illumination versus coherent-state target detection]{Quantum illumination versus coherent-state target detection}
\author{Jeffrey H Shapiro and Seth Lloyd}
\address{Research Laboratory of Electronics,
Massachusetts Institute of Technology, Cambridge, MA 02139, USA}
\ead{jhs@mit.edu}

\begin{abstract}
Lloyd \cite{Lloyd} proved that a large performance gain accrues from use of entanglement in single-photon target detection within a lossy, noisy environment when compared to what can be achieved with unentangled single-photon states.  We show that the performance of Lloyd's ``quantum illumination'' system is, at best, equal to that of a coherent-state transmitter, and may be substantially worse.  Nevertheless, as shown in \cite{PRL}, quantum illumination may offer a significant performance gain when operation is \em not\/\rm\ limited to the single-photon regime.  
\end{abstract}

\maketitle 
Consider a quantum-illumination radar system in which, on each transmission, one photon from a maximally-entangled pair interrogates a region of space in which a target might be located, while the other photon is retained at the receiver for use in a joint measurement  with what is received from that region.  Lloyd showed \cite{Lloyd} that such a  transmitter affords dramatically improved photodetection sensitivity in an entanglement-breaking scenario in comparison to what is achieved with unentangled single-photon states.  His quantum Chernoff bound on the error probability---when the optimum quantum measurement over $N$ repeated entangled-state transmissions is employed---can be stated as follows,
\begin{equation}
\Pr(e)_{\rm QI} \le \left\{\begin{array}{ll}
e^{-N\kappa}/2, & \mbox{for $\kappa \ll 1$, $MN_B \ll 1$, and $\kappa \gg N_B/M$}; \\ 
& \mbox{the ``good'' regime}\\[.12in]
e^{-N\kappa^2M/8N_B}/2, & \mbox{for $\kappa \ll N_B/M \ll MN_B \ll 1$};\\
& \mbox{the ``bad'' regime.}
\end{array}\right.
\label{LloydQI}
\end{equation}
Here, $\kappa$ is the transmitter-to-receiver coupling when the target is present, $N_B$ is the average number of received background photons per mode, and $M \gg 1$ is the number of temporal modes over which the transmitter state is entangled.  The distinction between ``good'' and ``bad'' regimes is that performance in the former is independent of the background noise, whereas the performance in the latter is dominated by that noise.  

The quantum illumination performance in (\ref{LloydQI}) is substantially better than the quantum Chernoff bound Lloyd found for $N$ repeated transmissions of a single-photon pure state, 
\begin{equation}
\Pr(e)_{\rm SP} \le \left\{\begin{array}{ll}
e^{-N\kappa}/2, & \mbox{for $\kappa \ll 1$ and $\kappa \gg N_B \ll MN_B \ll 1$};\\
& \mbox{the ``good'' regime}\\[.12in]
e^{-N\kappa^2/8N_B}/2, & \mbox{for $\kappa \ll N_B \ll MN_B \ll 1$};\\
& \mbox{the ``bad'' regime,}\end{array}\right.
\label{LloydSingle}
\end{equation}
where $M$-mode photodetection has been assumed, as in the entangled case.  Comparing 
(\ref{LloydQI}) and (\ref{LloydSingle}) shows that when both systems are in their ``good'' regimes, they achieve identical performance.  However, the ``good'' regime for quantum illumination extends to $M$-times higher background levels than does that for unentangled single-photon transmission.  Moreover, when both systems are in their ``bad'' regimes, quantum illumination achieves an error-probability exponent that is $M$ times higher than that for unentangled single-photon transmission.  

Lloyd's results were predicated upon the assumption of single-photon operation.  For any given transmission he assumed that at most one photon was received, be it from target return or background light.  He also indicated that a full Gaussian-state analysis of quantum illumination in the $\kappa \ll 1$ regime would lift the preceding restriction.  That analysis has now been performed by Tan \em et al\/\rm.\@ \cite{PRL}.  Their work, as we will now show, demonstrates a significant difference from what Lloyd found for the single-photon limit.  Before turning to that demonstration, it is important to note that Chernoff bounds are exponentially tight.  Specifically, if 
\begin{equation}
\Pr(e) \le e^{-N{\cal{E}}}/2
\end{equation}
is the Chernoff bound on the error probability achieved with $N$ repeated uses of a particular transmitter state, then
\begin{equation}
\lim_{N\rightarrow \infty}[\ln(\Pr(e))/N] = -{\cal{E}},
\end{equation}
so that system performance is well characterized by the Chernoff bound error-probability exponent ${\cal{E}}$.  

For a transmitter that emits a coherent state with $N$ photons on average---e.g., by $N$ repeated transmissions of a coherent state with unity average photon number---the quantum Chernoff bound from Tan \em et al\/\rm.\@ is
\begin{equation}
\Pr(e)_{\rm CS} \le e^{-N\kappa(\sqrt{N_B+1}-\sqrt{N_B})^2}/2 \approx
e^{-N\kappa}/2,\quad\mbox{for $N_B \ll 1$}.
\label{PeCS}
\end{equation}
This performance equals that of Lloyd's quantum illumination transmitter in that system's ``good'' regime,  and is \em superior\/\rm\ to that transmitter's performance in its ``bad'' regime.  Indeed, when (\ref{LloydQI}) is compared with (\ref{PeCS}) we see that the latter has no $\kappa$ restriction on its validity, i.e., $N_B \ll 1$ is enough to ensure that repeated transmission of coherent states with unity average photon number leads to minimum error probability target detection performance that is \em not\/\rm\ background-noise limited.  Thus, whereas Lloyd's quantum illumination system will fall prey to background noise---within its assumption of single-photon operation---when $\kappa \ll N_B/M \ll MN_B \ll 1$, no such thing happens for the coherent-state system.  

It is easy to understand the origin of the performance advantage afforded by $N$ repeated transmissions of a unity average photon-number coherent state.  All of these transmissions have been assumed to be phase coherent.  Hence they are equivalent to transmitting the coherent state $|\sqrt{N}\rangle$, with average photon number $N$, in a single mode that is the coherent superposition of the modes excited by each individual transmission.  Unity quantum efficiency photon counting (direct detection) on this super-mode will achieve error probability $e^{-N\kappa}/2$, when $N\kappa \gg N_B \ll 1$.  The optimum quantum receiver for the coherent-state system is not photon counting, but, as shown in (\ref{PeCS}), it is always in the ``good'' regime when $N_B \ll1$.   

Another way to exhibit the role played by coherence in achieving the performance given in (\ref{PeCS}) is to consider a receiver that eschews this coherence.  Suppose that on each of the $N$ coherent-state transmissions we employ a minimum error probability receiver to decide---based on the return from that transmission alone---whether or not the target is present.  When $\kappa \gg N_B \ll 1$, the error probability for this single-transmission receiver is given by
\begin{equation}
\Pr(e)_{{\rm CS}_1} \approx \frac{1-\sqrt{1-e^{-\kappa}}}{2} \approx \frac{1-\sqrt{\kappa}}{2},\,\mbox{ for $\kappa \ll 1$,}
\end{equation}
where we have used the background-free error probability of the optimum receiver in the first approximation \cite{Helstrom}.  Now suppose that the \em decisions\/\rm\ made on each of the $N$ individual transmissions are combined, by majority vote, to determine a final decision as to target absence or presence.  This majority-vote receiver ignores the coherence between the different coherent-state transmissions.  The Chernoff bound on its error probability is
\begin{equation}
\Pr(e)_{\rm MV} \le \frac{[2\sqrt{p(1-p)]}^N}{2},\, \mbox{ where $p \equiv \Pr(e)_{{\rm CS}_1}$.}
\end{equation}
Making use of the $\kappa \ll 1$ approximation for $p$ we can reduce this Chernoff bound to
\begin{equation}
\Pr(e)_{\rm MV} \le e^{-N\kappa/2}/2,
\end{equation}
which has an error-probability exponent that is a factor of two worse than what is obtained with coherent processing of the $N$ coherent-state transmissions.  

The single-shot receiver needed to achieve $\Pr(e)_{{\rm CS}_1}$ is a complicated photodetection feedback system \cite{Dolinar}.  Thus it is of interest to exhibit an alternative, for use with the coherent-state transmitter, whose performance  may exceed that of Lloyd's quantum illumination system.  Suppose that a coherent-state transmitter is used in conjunction with unity  quantum-efficiency homodyne detection.  The Chernoff bound for this system is
\begin{equation}
\Pr(e)_{\rm hom} \le e^{-N\kappa/(4N_B + 2)}/2 \approx e^{-N\kappa/2}/2,\, \mbox{ for $N_B \ll 1$,}
\end{equation}
so that the performance of this conventional (coherent-state transmitter, homodyne-detection receiver) laser radar is a factor of two worse in error-probability exponent than Lloyd's quantum illumination system when that system is in its ``good'' regime.   The conventional system, however, performs far better than Lloyd's quantum illumination system when the latter is in its ``bad'' regime.   

In conclusion, the full Gaussian-state analysis suggests that Lloyd's quantum illumination is unlikely to substantially improve radar performance in the low-noise regime wherein $N_B \ll 1$.  Tan {\em et al}.\@ \cite{PRL}, however, showed that a factor-of-four error-probability exponent improvement, over coherent-state transmission, \em is\/\rm\ achieved with a spontaneous parametric downconverter entangled-photon source and the optimum quantum measurement in the lossy ($\kappa \ll 1$), noisy ($N_B \gg 1$), low-brightness ($N/M \ll 1$) scenario.

\ack
This research was supported by the W. M. Keck Foundation Center for Extreme Quantum Information Theory and the DARPA Quantum Sensors Program.\\

\end{document}